Łukasz Kapłon*, Andrzej Kochanowski, Marcin Molenda, Paweł Moskal, Anna Wieczorek, Tomasz Bednarski, Piotr Białas, Eryk Czerwiński, Grzegorz Korcyl, Jakub Kowal, Paweł Kowalski, Tomasz Kozik, Wojciech Krzemień, Szymon Niedźwiecki, Marek Pałka, Monika Pawlik, Lech Raczyński, Zbigniew Rudy, Piotr Salabura, Neha Gupta-Sharma, Michał Silarski, Artur Słomski, Jerzy Smyrski, Adam Strzelecki, Wojciech Wiślicki, Marcin Zieliński and Natalia Zoń

# Plastic scintillators for positron emission tomography obtained by the bulk polymerization method

**Abstract:** This paper describes three methods regarding the production of plastic scintillators. One method appears to be suitable for the manufacturing of plastic scintillator, revealing properties which fulfill the requirements of novel positron emission tomography scanners based on plastic scintillators. The key parameters of the manufacturing process are determined and discussed.

*Corresponding author: Łukasz Kapłon, Faculty of Physics, Institute of Physics, Astronomy and Applied Computer Science, Jagiellonian University, ul. Reymonta 4, 30-059 Krakow, Poland, Tel.: +48-12-6635736, E-mail: lukasz.kaplon@uj.edu.pl; Institute of Metallurgy and Materials Science of Polish Academy of Sciences, Krakow, Poland
**Andrzej Kochanowski and Marcin Molenda:** Faculty of Chemistry, Jagiellonian University, Krakow, Poland
**Paweł Moskal, Tomasz Bednarski, Piotr Białas, Eryk Czerwiński, Grzegorz Korcyl, Jakub Kowal, Tomasz Kozik, Wojciech Krzemień, Szymon Niedźwiecki, Marek Pałka, Monika Pawlik, Zbigniew Rudy, Piotr Salabura, Neha Gupta-Sharma, Michał Silarski, Artur Słomski, Jerzy Smyrski, Adam Strzelecki, Marcin Zieliński and Natalia Zoń:** Faculty of Physics, Astronomy and Applied Computer Science, Jagiellonian University, Krakow, Poland
**Anna Wieczorek:** Faculty of Physics, Astronomy and Applied Computer Science, Jagiellonian University, Krakow, Poland; and Institute of Metallurgy and Materials Science of Polish Academy of Sciences, Krakow, Poland
**Paweł Kowalski, Lech Raczyński and Wojciech Wiślicki:** Swierk Computing Centre, National Centre for Nuclear Research, Otwock-Swierk, Poland

# Introduction

Commercial positron emission tomography (PET) scanners are based on inorganic crystals such as bismuth germanium oxide, lutetium oxyorthosilicate, and lutetium yttrium oxyorthosilicate [1]. However, two new solutions were recently proposed for detecting gamma quanta originating from positron annihilation in the human body [2]. The first method involves the use of plastic scintillator strips arranged in the form of a barrel [3], called "strip PET", and the second proposed method is based on plastic scintillator plates combined with arrays of photomultipliers, called "matrix PET" [4]. Both solutions are based on polymer scintillators with special fluorescent additives. The main difference in the detection principle is the utilization of time properties of the scintillators instead of signal amplitudes. Plastic scintillators have at least one order of magnitude shorter decay time than inorganic crystals. This feature allows the improvement of the resolution of time-of-flight (TOF) determination in a stripPET apparatus, which is presently being developed by the Jagiellonian-PET collaboration.

This paper describes the production methods of plastic scintillators found in the literature and adapted for requirements in novel PET scanner applications. The key parameters of the manufacturing process are determined and discussed.

## Scintillator requirements for novel PET scanners

Plastic scintillators are suitable for application in TOF detectors due to their short response time and the possibility of production in various shapes and sizes. The achievable time resolution of a PET scanner depends on the decay and rise time of light signals produced in scintillators and on the amount of light reaching the photomultipliers. The decay time of a typical plastic scintillator ranges from 1.4 to 2.4 ns and light output amounts to approximately 10,000 photons/MeV of absorbed gamma radiation [5]. A large attenuation length of up to 400 cm allows the transportation of light from the center to the edges of the scintillator strips with small losses. The maximum of emission spectra is observed at around 420 nm and this value matches well with the quantum efficiency of typical photomultiplier tubes.

**Table 1** Physical properties of polymers for scintillator production [8].

| Name | Abbreviation | Density, g/cm$^3$ | Glass-transition temperature, °C | Refractive index |
|---|---|---|---|---|
| Polystyrene | PS | 1.04-1.06 | 100 | 1.59 |
| Polyvinyltoluene | PVT | 1.02 | 93-118 | 1.58 |

Commercially available plastic scintillators are mainly made of polystyrene or polyvinyltoluene with fluorescent additives [6, 7]. The properties of plastic scintillators depend not only on fluorescent compounds but also arise from properties of polymer base. The quality of the polymer affects the attenuation length coefficient and light output of the resulting scintillator. Some physical properties of polymers are presented in Table 1. All components are required to be purified prior to the manufacturing process because impurities attenuate light and quench fluorescence.

The light output of plastic scintillators depends on the average molecular weight of the polymer [9]. For example, a polystyrene value above 100,000 units is sufficient to obtain acceptable properties of a scintillator. With a value under 100,000 units, the light output is increasing with the increasing average molecular weight of the polymer. This molecular weight may be obtained through the manufacturing process, called batch cell casting. This technique involves polymerization of the solution of liquid monomer with fluorescent additives without polymerization initiators. The process is based on thermal radical polymerization that uses heat of the polymerization reaction and heat provided from electrical heaters in the furnace. The obtained scintillator reveals good optical properties and looks like organic glass, that is, it has an amorphous structure. This scintillator is easily mechanically machined to a desirable shape and dimension of up to a few meters (dimensions are limited by the size of the furnace).

## Industrial methods of plastic scintillator production

There are three main manufacturing methods of polymers realized at the industrial scale that are applied to the production of plastic scintillators. The best results are obtained in batch cell casting and this method is commonly used to make commercial scintillators. The other two methods concerning the production of polymers at the industrial scale are the injection molding technique and the extrusion technique. Both involve mixing a solid polymer with additions and use expensive equipment.

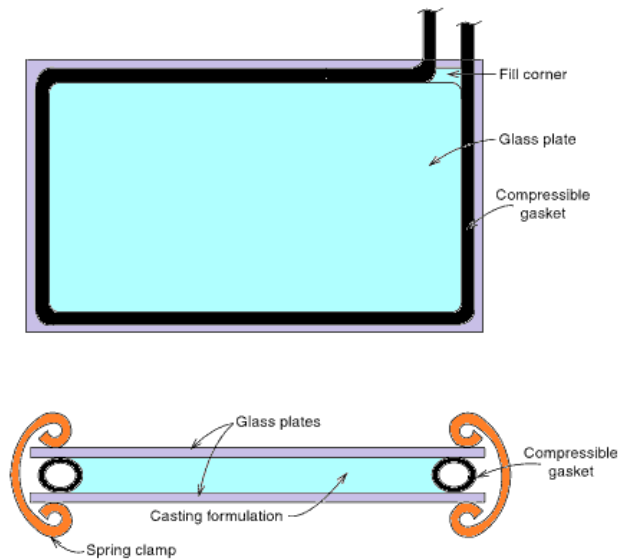

**Figure 1** Face view (top) and edge view (bottom) of conventional cell casting mold configuration. Scheme adapted from [10].

Batch cell casting is a process where a liquid monomer with dissolved dopants is poured into a mold and heated to rigid solid plastic (see Figure 1). In the first step, the monomer is purified on activated alumina sorbent. The sorbent removes impurities such as inhibitor and water from the monomer. Dissolved oxygen and other gases are removed from the solution by degassing under reduced pressure. Before pouring the solution into the mold, the surface of the mold is treated with a solution of dichlorodimethylsilane in chloroform. This procedure is called silanization and allows the prevention of adherence of the polymer sample to the glass mold by formation of the antiadhesive layer. The mold is then placed in the furnace under a heating cycle that takes place for approximately 5 days. After a few days the heating scintillator is cooled, annealed, and mechanically cut and polished. In this way, it is possible to obtain scintillators in shapes of blocks, plates, sheets, rods and bars. More complicated profiles may be manufactured by lathe work.

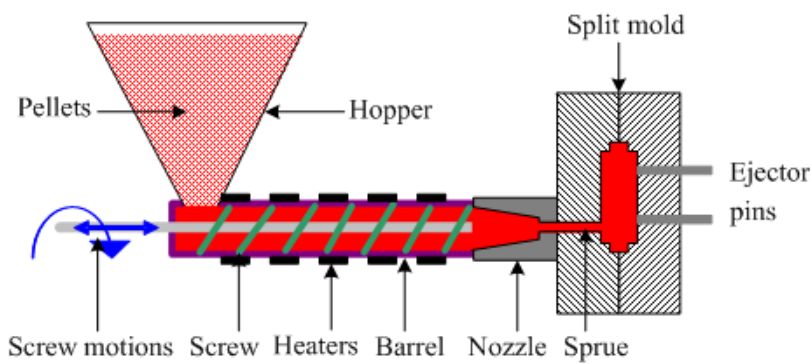

**Figure 2** Schematic of typical screw-injection molding machine. Adapted from [12].

The injection molding technique is widely used in industry. Most plastic goods and packages are produced with this technique. Application of the injection molding technique for mass production of scintillation tiles was first developed in the early 1980s [11]. With this process, optically transparent granulated polystyrene is mixed with scintillation dopants. Then the mixture is loaded into a molding machine hopper, where it is continuously directed into a heated screw cylinder while being mixed (the scheme in Figure 2). At the exit of the cylinder the temperature reaches approximately 200°C and the melted polystyrene accumulates in its nozzle. As it becomes full, an injection into the mold starts. It lasts for approximately 3 s at a pressure of approximately 700 atm. After mold cooling up to approximately 50°C, it opens and the tile is taken away. The whole cycle lasts for <2 min/tile. The production rate is high and the cost is a small fraction of the cost of a commercial scintillator. In addition, no secondary mechanical operation is needed for the final product. Scintillators produced with the injection molding technique have inferior light yields and poorer optical properties compared with cast scintillators. It is related to the speed of the cooling rate of the scintillator being too high, which results in optical heterogeneity. Moreover, a very high temperature during manufacturing causes degradation of the polymer.

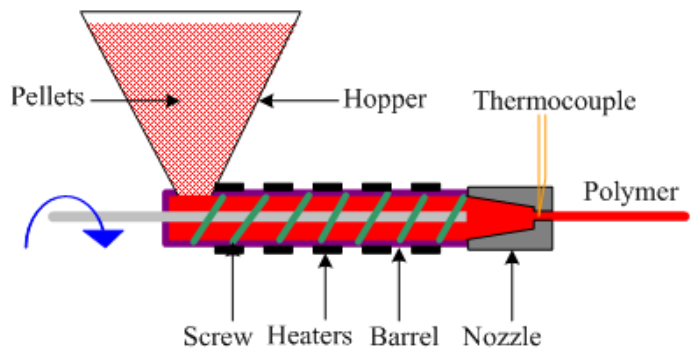

**Figure 3** Schematic of typical screw-extrusion machine. Adapted from [12].

The extrusion technique was first applied to plastic scintillator detectors in 1980 [13] and produced a polystyrene-based scintillator with good light yield, but demonstrated a relatively poor attenuation length. Twenty years later, another attempt was performed in order to make better scintillators with lower costs [14]. This method is a continuous in-line compounding and extrusion process (presented in Figure 3). Similar to the injection molding technique, polymer pellets are mixed with fluorescence dopants and loaded to a molding machine hopper with heaters and mixing steps. At the end of the machine there is a profile dye instead of mold. The extruder function is to melt, mix, and deliver the scintillator material to the profile dye. All steps are done under argon gas to prevent degradation of the polymer. After extrusion, the material immediately enters the vacuum-sizing tool mounted in a long chilled water tank. In this vacuum-sizing tank, differential pressure draws the semi-molten material to final dimensions. The material is further cooled in a second water-chilled section. The resulting scintillator has a shape of a long strip with a cross-section of the dye.

## Laboratory tests

For laboratory work, the batch cell casting technology is used to obtain samples of plastic scintillators for further investigation. For this purpose, a tube furnace with automatic and autonomic control of the process parameters is used. The furnace has four independent heating zones, seven steps of time and temperature per cycle. The maximum achievable temperature is up to 260°C and is controlled with an accuracy of 1°C. The following parameters were controlled: temperature, time, heating and cooling rate in each zone. All parameters are programmable from a computer and can be saved on-line to a personal computer with the thermal history of each zone.

For research and development purposes, styrene as a monomer was chosen to produce polystyrene scintillators. A styrene monomer has low cost, is commercially available, and scintillators obtained on this basis have almost the same properties as for polyvinyltoluene. To obtain high quality polymers, polymerization initiators, such as azobisisobutyronitrile or benzoyl peroxide, have to be avoided. Both initiators cause rapid conversion from monomer to polymer. Additionally, benzoyl peroxide is responsible for yellowing of polystyrene during polymerization. Also, the use of crosslinking the monomer divinylbenzene has adverse effects, for example, the resulting copolymer styrene-divinylbenzene shrinks too much in the mold and degrades the shape of scintillators. Styrene was purified on activated alumina sorbent balls. As a mold, glass ampoules 25 mm in diameter were used and were treated with a solution of dichlorodimethylsilane in chloroform to prevent stick scintillator samples to glass.

Studies on the polymerization process were performed to evaluate the optimal time and temperature cycle conditions. A sample temperature cycle is shown in Figure 4. The first step is related to heating from room temperature to 140°C in 5–10 h. In this part, polymerization starts and for small samples requires a rapid heating ratio owing to vacuum bubbles forming. If the temperature is still below the glass-transition temperature of polystyrene (100°C) and the reaction takes place in viscous solution of the polymer in monomer, then vacuum bubbles occur.

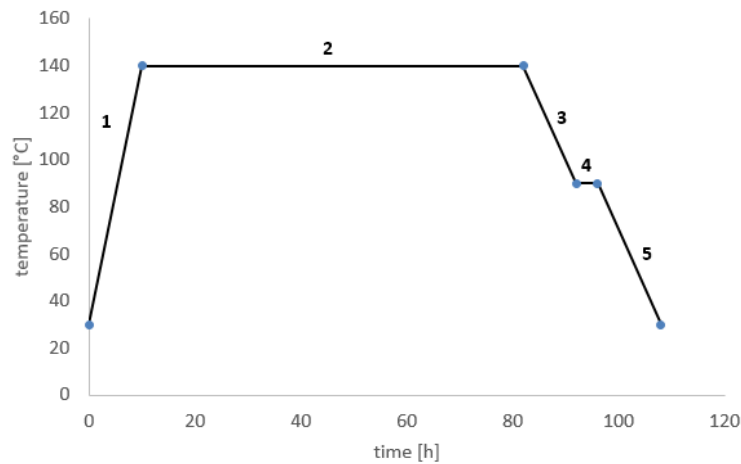

**Figure 4** Temperature cycle for polymerization of styrene.

The second step involves polymerization at 140°C for approximately 3 days and ensures complete conversion from monomer to polymer. Following the main polymerization step, slow cooling is applied (third step) under glass-transition temperature. Further annealing (fourth step) of polymer scintillators at 90°C for 4 h is required because rapid cooling can crack the polymer sample due to internal stress from shrinking with decreasing temperature. Finally, the fifth step is related to cooling to room temperature.

# Conclusions and discussion

Three methods of the manufacturing process of plastic scintillators have been described. A comparison of the described methods is presented in Table 2. Only one technique, batch cell casting, is commonly applied to fabricate plastic scintillators in industry and in research and development laboratories. This method seems to be suitable for the manufacturing of plastic scintillators, revealing properties which will fulfill the requirements of novel PET scanner applications.

**Table 2** Comparison of industrial methods for production of plastic scintillators.

| Technique | Advantages | Drawbacks |
|---|---|---|
| Cell casting | Best polymer quality and scintillator performance, smooth scintillator surfaces from mold, easily mechanical machined | High manufacturing costs, time-consuming process, complicated reaction preparation |
| Injection molding | Low cost, ability to produce complicated shapes | Optical heterogeneities, mechanical stresses inside the polymer, much lower scintillator performance |
| Extrusion | Low cost, ability to produce long strips with any cross-section | Optical heterogeneities, mechanical stresses inside the polymer, much lower scintillator performance |

**Acknowledgments:** We acknowledge technical and administrative support from M. Adamczyk, T. Gucwa-Rys, A. Heczko, M. Kajetanowicz, G. Konopka-Cupiał, J. Majewski, W. Migdał, A. Misiak, and financial support by the Polish National Center for Development and Research through grant INNOTECH-K1/IN1/64/159174/NCBR/12, the Foundation for Polish Science through International PhD Projects (MPD) and The EU and Ministry of Science and Higher Education Grant No. POIG.02.03.00-161 00-013/09.

**Conflict of interest statement**

**Authors' conflict of interest disclosure:** The authors stated that there are no conflicts of interest regarding the publication of this article. Research funding played no role in the study design; in the collection, analysis, and interpretation of data; in the writing of the report; or in the decision to submit the report for publication.

**Research funding:** None declared.
**Employment or leadership:** None declared.

**Honorarium:** None declared.

Received October 29, 2013; accepted January 30, 2014; previously published online February 22, 2014